\documentstyle[12pt]{article}
\def\singlespace {\smallskipamount=3.75pt plus1pt minus1pt
                  \medskipamount=7.5pt plus2pt minus2pt
                  \bigskipamount=15pt plus4pt minus4pt
                  \normalbaselineskip=15pt plus0pt minus0pt
                  \normallineskip=1pt
                  \normallineskiplimit=0pt
                  \jot=3.75pt
                  {\def\smallskip {\vskip\smallskipamount}}
                  {\def\medskip   {\vskip\medskipamount}}
                  {\def\bigskip   {\vskip\bigskipamount}}
                  {\setbox\strutbox=\hbox{\vrule
                    height10.5pt depth4.5pt width 0pt}}
                  \parskip 7.5pt
                  \normalbaselines}
\def\middlespace {\smallskipamount=5.825pt plus1.5pt minus1.5pt
                  \medskipamount=11.25pt plus3pt minus3pt
                  \bigskipamount=22.5pt plus6pt minus6pt
                  \normalbaselineskip=22.5pt plus0pt minus0pt
                  \normallineskip=1pt
                  \normallineskiplimit=0pt
                  \jot=5.825pt
                  {\def\smallskip {\vskip\smallskipamount}}
                  {\def\medskip   {\vskip\medskipamount}}
                  {\def\bigskip   {\vskip\bigskipamount}}
                  {\setbox\strutbox=\hbox{\vrule
                    height15.75pt depth6.75pt width 0pt}}
                  \parskip 7.25pt
                  \normalbaselines}
\def\dblspc {\smallskipamount=7.5pt plus2pt minus2pt
                  \medskipamount=15pt plus4pt minus4pt
                  \bigskipamount=30pt plus8pt minus8pt
                  \normalbaselineskip=30pt plus0pt minus0pt
                  \normallineskip=2pt
                  \normallineskiplimit=0pt
                  \jot=7.5pt
                  {\def\smallskip {\vskip\smallskipamount}}
                  {\def\medskip   {\vskip\medskipamount}}
                  {\def\bigskip   {\vskip\bigskipamount}}
                  {\setbox\strutbox=\hbox{\vrule
                    height21.0pt depth9.0pt width 0pt}}
                  \parskip 15.0pt
                  \normalbaselines}

\def\rar{\rightarrow}
\def\be{\begin{equation}}

\def\j-{\J_-}

\def\ee{\end{equation}}

\def\bearr{\begin{eqnarray}}
\def\bearrs{\begin{eqnarray*}}
\def\eearr{\end{eqnarray}}
\def\eearrs{\end{eqnarray*}}
\def\barr{\begin{array}}
\def\earr{\end{array}}

\def\non\non{\nonumber}
\def\nn8{\nonumber\\[15pt]}
\def\l{\left}
\def\r{\right}
\def\un{\underline}
\def\ve{\varepsilon}
\def\f{\frac}

\begin{document}
\thispagestyle{empty}
\begin{center}
{\huge No birefringence in Einstein's gravity}\\[12pt]
{\bf S. Mohanty and A.R. Prasanna}\\
Physical Research Laboratory\\
Ahmedabad 380 009\\
India\\
\end{center}

\dblspc
{\bf Einstein's theory predicts that, massive test particles with
non-zero spin or angular momentum, in an external gravitation
field, follow geodesics which  depend upon the orientation of
the spin-angular momentum [1]. It has been claimed [2] that such
an effect also holds for photons i.e. photons with different
helicities follow different geodesics in the gravitational field
of a rotating body. If such an effect were to exist it would
result in a polarisation dependent deflection of light passing
in the vicinity of the sun [3] or a polarisation dependent time
delay of pulsar signals [4]. We show here that contrary to
earlier claims [2], in Einstein's gravity there is no
birefringence and photons follow null geodesics irrespective of
polarisation. Thus if gravitational birefringence is ever
observed experimentally  it would be a signal of new physics
beyond Einstein's gravity [5]}.\\

Maxwell's equations in curved space follow from the equation of motion
\be
\nabla_\mu \sqrt{g} F^{\mu \nu} = 0
\ee
and the Bianchi identity
\be
\nabla_\mu \tilde{F}_{\mu \nu} = 0
\ee
where $\tilde{F}_{\mu \nu} = \f{1}{2} \varepsilon_{\mu \nu
\alpha \beta} F_{\alpha \beta}$ is the dual of the
electromagnetic field tensor $F_{\mu \nu}$ and the derivatives
in (1) and (2) are covariant w.r.t. the background metric.\\

In a Cartesian frame the field tensor can be decomposed in terms
of the electric and magnetic fields as $F_{\mu \nu} \rar \l(
\mbox{\boldmath E}.
\mbox{\boldmath B} \r)$ and $\sqrt{-g} F^{\mu \nu} \rar (- \mbox{\boldmath D},
\mbox{\boldmath H} )$. The
electromagnetic wave equation in curved space can be obtained by
introducing the vector $\mbox{\boldmath F}^\pm = \mbox{\boldmath
E} \pm i \mbox{\boldmath H}$. Two of the Maxwell's
equations from (1) and (2) can be written as
\be
{\bf \nabla} \times \mbox{\boldmath F}^\pm = \pm i
\omega \ve \mbox{\boldmath F}^\pm + i \omega \mbox{\boldmath G} \times
\mbox{\boldmath F}^\pm
\ee
\be
 \nabla \cdot \mbox{\boldmath F}^\pm = i \omega \mbox{\boldmath
G} \cdot \mbox{\boldmath F}^\pm
\ee
where $\varepsilon$ is a $3\times3$ matrix with components
$\varepsilon_{ij} = \left( g^{ij}/g_{oo} \right) \sqrt{-g}$ and
$\mbox{\boldmath G}$ is a three vector with components $G_i = -
g_{oi}/g_{oo}$, and
we have assumed that the fields have a harmonic time
dependence with frequency $\omega$. The wave equation for
$\mbox{\boldmath F}^\pm$ is obtained by taking the curl of (3) and (4) to give
\be
\l[ - {\bf \nabla}^2 - \omega^2 \ve^2 + 2 i \omega
\mbox{\boldmath G} \cdot {\bf \nabla} \r]
\mbox{\boldmath F}^\pm = 0
\ee
 From (5) it is clear that the wave equations for both the right
circular polarisation $\l( \mbox{\boldmath F}^\pm \r)$ and the left circular
polarisation $\l( \mbox{\boldmath F}^- \r)$ modes are identical. This is not a
result of the limitation of the geometrical optics limit as
claimed in [2] but a consequence of the exact wave equation is
curved space (5).  Since the derivatives in (5) are covariant
w.r.t. the background metric, the direction of polarisation of
$\mbox{\boldmath F}^\pm$ undergoes a Fermi-Walker transport in the course of
the
propagation of the wave along the null geodesic, which is
interpreted as a gravitational Faraday rotation [6].
In [2] the author solves the first order Maxwell's equation (3)
to show that the wavenumbers $k^\pm$ of the modes
$\mbox{\boldmath F}^\pm$ differ by a constant proportional to
the rotational parameter of the metric. This however means that
the two circularly polarised modes propagate with different
phase velocities but the same group velocity - therefore
Einstein's gravity causes a Faraday rotation but no birefringence.
If such birefringent effects
like  polarisation dependent bending of light by the
Sun or time delay of pulsar signals are observed, they
will signal new physics
beyond Einstein's gravity [6].\\[20pt]

\noindent
{\bf References}
\begin{itemize}
\item[[1]] Papapetrou A., Proc. R. Soc. London \un{A209},
248-258 (1951).
\item[[2]] Mashoon, B., Nature \un{250}, 316-317 (1974);\\
{\it ibid}, Phys. Rev. \un{D7}, 2807-2814 (1973).
\item[[3]] Harwit, M., et. al., Nature \un{249}, 230-233 (1974).
\item[[4]] Cordes, J.M. and Stinebring, D.R., Ap.L. Lett.
\un{277}, 53 (1984);\\
Loseco, J.M., et.al., Phy. Lett. \un{A138}, 5 (1989);\\
Klien, J.R. and Thorsett, S.E., Phy. Lett. \un{A145}, 79 (1990).
\item[[5]] Prasanna, A.R. and Mohanty, S., `` Astrophysical
Tests of CP Violations by Gravity", PRL Preprint.
\item[[6]] Strotskii, G.V., Soviet Phys. Dokl. \un{2}, 226-229 (1957).
\end{itemize}
\end{document}